\documentclass[aps, amsmath, floats, floatfix, twocolumn,
superscriptaddress, nofootinbib, showpacs]{revtex4-1} 

\usepackage{graphicx}
\usepackage{color}
\usepackage{soul}
\usepackage{url}
\usepackage{bm}         
\usepackage{times}
\usepackage{dcolumn}
\usepackage{bm}
\usepackage{epsf}
\usepackage{amssymb}

\newcommand{\beq}{\begin{equation}}
\newcommand{\eeq}{\end{equation}}
\newcommand{\beqn}{\begin{eqnarray}}
\newcommand{\eeqn}{\end{eqnarray}}

\newcommand{\UNH}{\affiliation {Department of Physics, University of New Hampshire, 9 Library Way, Durham NH 03824, USA}}

\begin{document}

\title{A brief overview of black hole-neutron star mergers} 

\author{F. Foucart}\UNH

\begin{abstract}

Of the three main types of binaries detectable through ground-based gravitational wave observations, black hole-neutron star (BHNS) mergers remain the most elusive. While candidates BHNS exist in the triggers released during the third observing run of the Advanced LIGO/Virgo collaboration, no detection has been confirmed so far. As for binary neutron star systems, BHNS binaries allow us to explore a wide range of physical processes, including the neutron star equation of state, nucleosynthesis, stellar evolution, high-energy astrophysics, and the expansion of the Universe. Here, we review some of the main features of BHNS systems: the distinction between disrupting and non-disrupting binaries, the types of outflows that BHNS mergers can produce, and the information that can be extracted from the observation of their gravitational wave and electromagnetic signals. We also emphasize that for the most likely binary parameters, BHNS mergers seem less likely to power electromagnetic signals than binary neutron star systems. Finally, we discuss some of the issues that still limit our ability to model and interpret electromagnetic signals from BHNS binaries. 
\end{abstract}

\maketitle

\section{Introduction}

The first observation by the LIGO-Virgo collaboration (LVC) of gravitational waves (GWs) coming from merging black holes~\citep[GW150914]{TheLIGOScientific:2016qqj} and neutron stars~\citep[GW170817]{TheLIGOScientific:2017qsa} made GW astrophysics a reality. Since then, the LVC has confirmed an additional 9 binary black holes (BBH)~\citep{LIGOScientific:2018mvr}, with dozens of other systems announced in public alerts.\footnote{Public alerts from the LVC are available online at https://gracedb.ligo.org/} BBH mergers were also discovered by an independent search pipeline used on publicly available LVC data~\citep{Venumadhav:2019lyq}. Most recently, an event that may have been a second binary neutron star (BNS) merger was reported by the LVC~\citep[GW190425]{Abbott:2020uma}\footnote{GW190425 has observed masses that could also plausibly be explained as a very low mass BHNS merger, if $\sim (2-3)M_\odot$ black holes exist.}.

BNS and black hole-neutron star (BHNS) systems play an especially interesting role in this new field of astrophysics. By observing neutron star mergers, we gather information about the equation of state of neutron stars~\citep{Flanagan2008,GW170817-NSRadius}, about the origin of heavy elements produced through r-process nucleosynthesis~\citep{Freiburghaus:1999no,Drout:2017ijr,2017Natur.551...67P}, and about the expansion rate of the Universe~\citep{Abbott:2017xzu,Hotokezaka:2018dfi}. 
Neutron star mergers also power at least a subset of short gamma-ray bursts (SGRBs)~\citep{2017ApJ...848L..13A}, as well as UV/optical/infrared {\it kilonovae}~\citep{Li:1998bw,Roberts2011,2017Sci...358.1556C,2017ApJ...848L..16S,2017ApJ...848L..18N,Cowperthwaite:2017dyu,2017Sci...358.1565E,2017ApJ...848L..19C,Villar:2017wcc}, and radio emission from ultra-relativistic jets and mildly relativistic outflows~\citep{Nakar:2011cw,Hotokezaka:2016,2018Natur.561..355M}.
The event rates of BNS and BHNS mergers remain however very uncertain~\citep{Abadie:2010cfa,belczynski:16,Abbott:2020uma}. Only two potential BNS mergers have been officially confirmed so far, and no BHNS mergers, even though a number of candidates can be found within the LVC's public alerts.

\begin{figure}[h!]
\begin{center}
\includegraphics[width=0.45\linewidth]{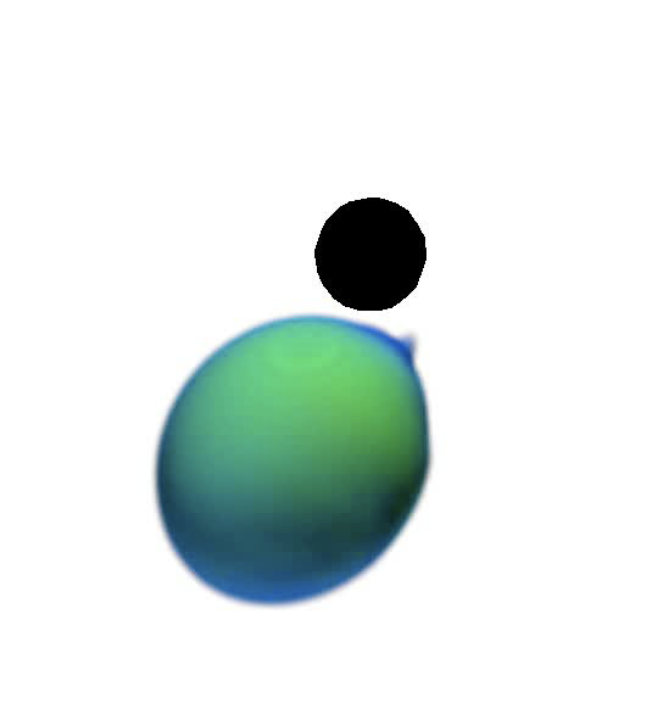}
\includegraphics[width=0.45\linewidth]{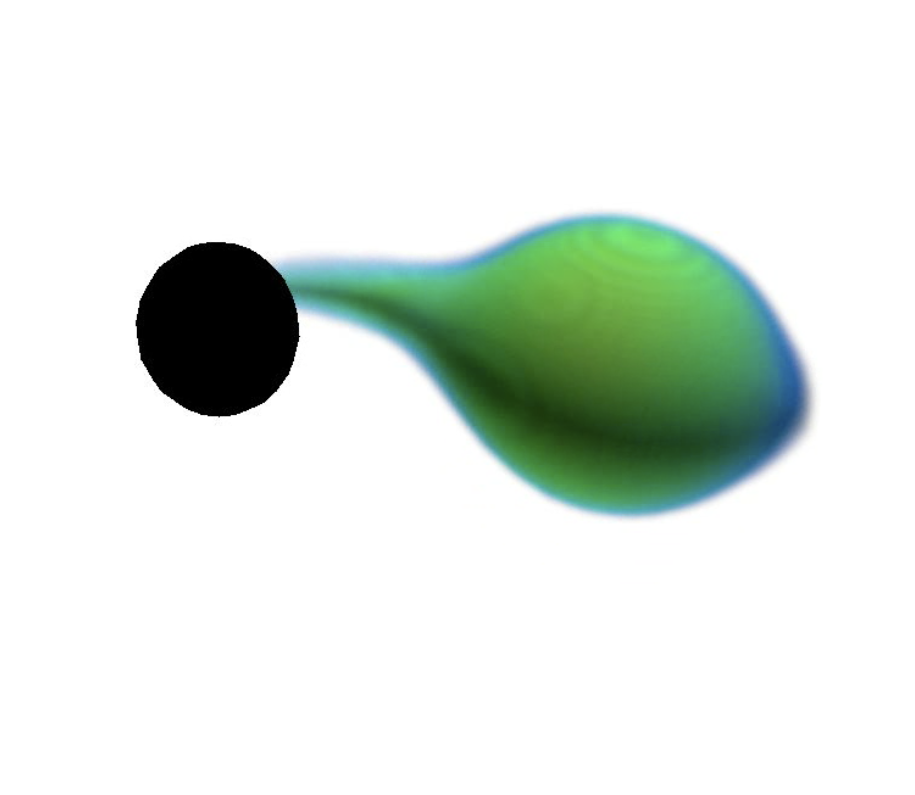}\\
\includegraphics[width=0.45\linewidth]{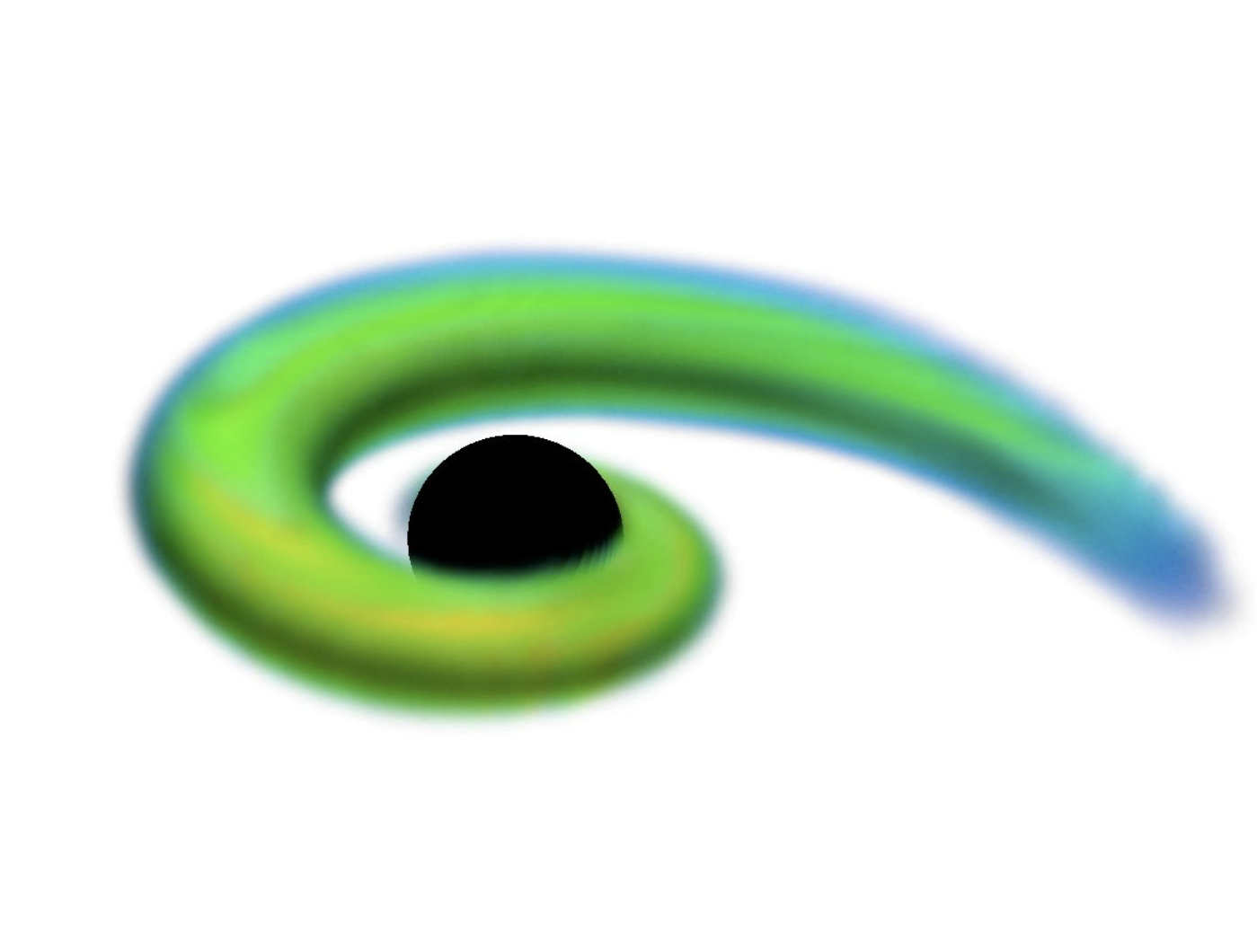}
\includegraphics[width=0.45\linewidth]{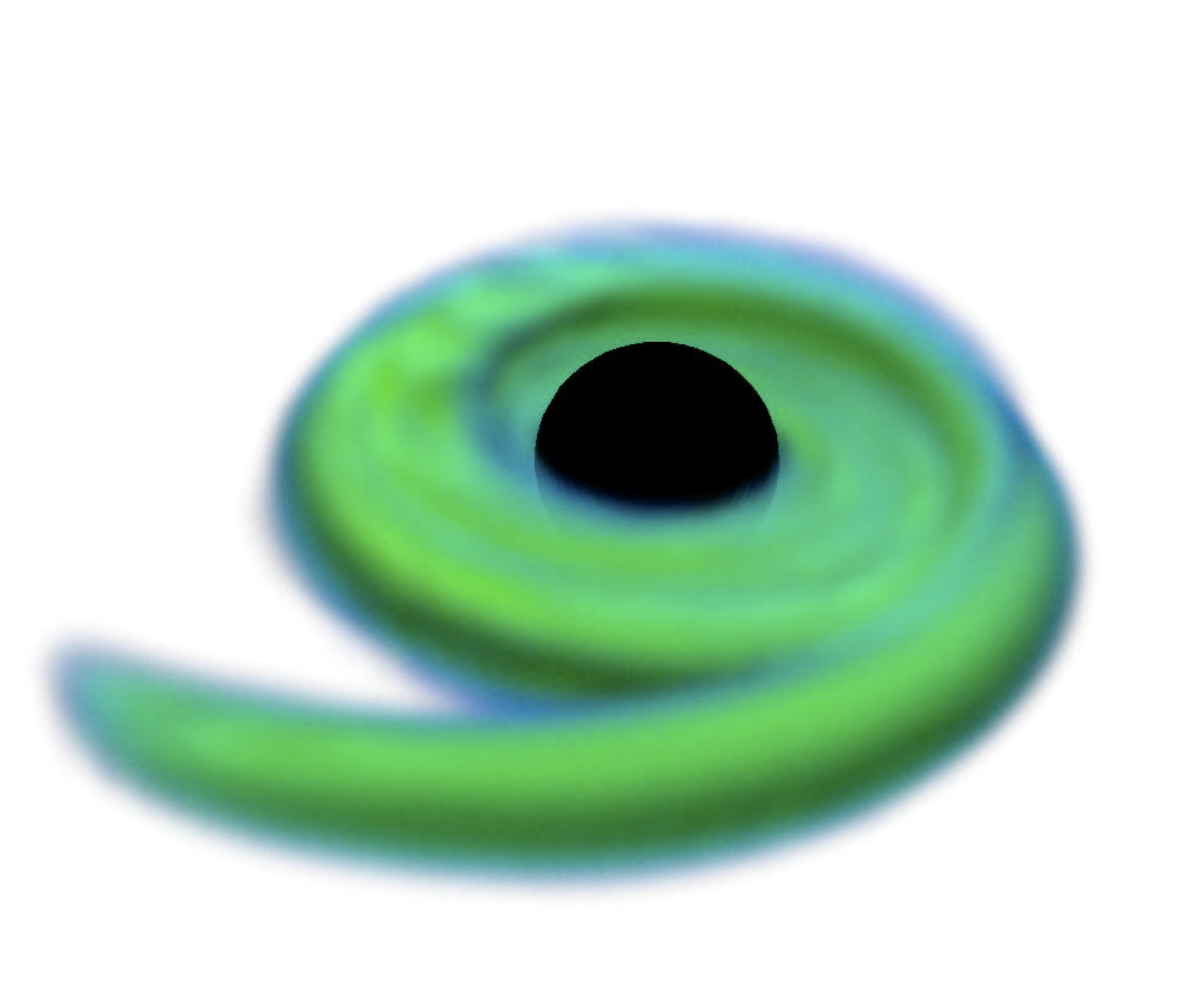}
\end{center}
\caption{Time evolution of a disrupting BHNS binary, including: onset of mass accretion (top left), unstable mass transfer (top right), evolution into a long tidal tail (bottom left), and circularization into an accretion disk (bottom right).}\label{fig:merger}
\end{figure}

The evolution of BHNS binaries can be divided into three main phases: a millions-of-years long inspiral during which the two objects slowly lose energy and angular momentum to GW emission; a merger phase lasting about one millisecond and resulting in either the tidal disruption of the neutron star (see Fig.~\ref{fig:merger}) or its plunge into the black hole; and, for disrupting systems only, a seconds-long post-merger phase during which more matter is ejected or accreted onto the black hole. These three phases happen on widely different timescales, and involve different physical processes and observable signals. In this manuscript, we will review each stage of a BHNS's evolution in turn, and discuss important properties of the associated GW and electromagnetic (EM) signals.

BHNS systems cover a high-dimensional and largely unconstrained parameter space. Our priors for the properties of black holes and neutron stars in BHNS binaries come from their observation in other types of binary systems or from theoretical considerations, and are accordingly quite uncertain. While most neutron stars observed in BNS systems have masses in the $[1.2-1.6]\,M_\odot$ range~\citep{Ozel2012}, more massive neutron stars exist, up to at least $\sim 2M_\odot$~\citep{Demorest:2010bx,Antoniadis:2013pzd}. Most galactic black holes have masses of $[5-15]\,M_\odot$\citep{Ozel:2010}, but black holes observed through GWs are often more massive~\citep{LIGOScientific:2018mvr}. Whether black holes can be formed within the ``mass gap'' between the most massive neutron stars and $\sim 5M_\odot$ also remains an important open question. The magnitude and orientation of black hole spins are unknown, and while most BHNS binaries are expected to have negligible eccentricities~\citep{PetersMathews1963}, eccentric BHNS binaries cannot entirely be ruled out and have evolutions very distinct from circular binaries~\citep{Stephens:2011as}. Obtaining reliable models for the observable signals powered by BHNS binaries across this vast parameter space can be difficult, yet the dependence of these signals in the properties of BHNS binaries is what allows us to extract valuable information from observations. In this review, we mostly consider circular binaries, leaving as free parameters the masses $M_{\rm NS,BH}$ of the compact objects, their dimensionless spins ${\bar \chi}_{\rm NS,BH}$, and the equation of state of dense nuclear matter, which sets the radius $R_{\rm NS}$ of a given neutron star. 

\section{Binary inspiral}

From an observational point of view, the millions of years of GW driven inspiral that eventually result in the merger of a BHNS binary constitute an extended dark age between the supernova explosion that created the neutron star and the bright GW and EM emissions that accompany the merger. Ground-based GW detectors such as LIGO and Virgo only become sensitive to BHNS binaries seconds to minutes before merger. Most of our efforts thus focus on understanding the very end of the inspiral. 
To first order, the GW-driven inspiral of BHNS binaries proceeds as for black holes of the same masses and spins. GW detectors are mostly sensitive to the chirp mass 
\beq
M_c = \frac{(M_{\rm NS}M_{\rm BH})^{3/5}}{(M_{\rm BH}+M_{\rm NS})^{1/5}}
\eeq
of a system, while individual mass measurements suffer from large statistical errors for all but the brightest events. As for BNS systems, the main observable effect of the finite size of neutron stars before merger is the acceleration of the GW-driven inspiral due to tides~\citep{Flanagan2008}: large neutron stars merge earlier than more compact stars. GW detectors are primarily sensitive to the resulting change in the phase of the GW signal. To first order, that change is linear in the dimensionless tidal deformability parameter~\citep{Hinderer2010}, defined for BHNS systems as
\beq
\tilde \Lambda = \frac{32}{39} \frac{M_{\rm NS}^4 (M_{\rm NS}+12M_{\rm BH})}{(M_{\rm NS}+M_{\rm BH})^5} \frac{k_2}{C_{\rm NS}^5}
\eeq
with $C_{\rm NS}=M_{\rm NS}G/(R_{\rm NS}c^2)$ the compaction of the neutron star, and $k_2$ its dimensionless $l=2$ Love number. Both $k_2$ and $R_{\rm NS}$ depend on the equation of state of nuclear matter inside the neutron star. Unfortunately, $\tilde \Lambda$ becomes very small when $M_{\rm BH} \gg M_{\rm NS}$. As a result, finite size effects in BHNS mergers are expected to be detectable only for close-by events involving low-mass black holes~\citep{Lackey:2013axa}. The usefulness of BHNS binaries for the determination of the neutron star equation of state largely depends on the event rate of BHNS mergers that involve low-mass black holes. The existence of black holes within the supposed ``mass gap'' would be particularly convenient in that respect.
For reference, recent results from the LVC~\citep{Abbott:2018exr}, NICER~\citep{Riley:2019yda,Miller:2019cac}, and joint analysis of both datasets~\citep{PhysRevD.101.123007,Raaijmakers:2019dks} find $R_{\rm NS}\sim (10.5-14.5)\,{\rm km}$, with variations due to the chosen astrophysical data, equation of state model and maximum NS mass.

Additionally, {\it it can be difficult to unequivocally determine that a given GW signal is powered by a BHNS merger}. In the absence of an EM signal, our main method to determine the nature of merging compact objects is to use their inferred masses, and to assume that any object below a fixed threshold mass is a neutron star. This clearly introduces an untested astrophysical prior in the interpretation of the data. 
It can also be difficult to determine whether a system is a high mass ratio BHNS system or a more symmetric BBH system with the same chirp mass~\citep{HannamEtAl:2013}.\footnote{If we allow for primordial black holes within the same mass range as neutron stars, low-mass BHNS mergers can also mimic BNS systems, even if we observe an EM counterpart~\citep{Hinderer:2018pei}. }
Furthermore, if black holes are commonly formed with large spins misaligned with the orbital angular momentum of the binary, BHNS binaries may experience significant orbital precession. As the GW templates currently used by detection pipelines do not take orbital precession into account, this could lead to the loss of a significant fraction ($\sim 30\%$) of BHNS systems~\citep{Harry:2013tca}, with an observational bias towards the detection of non-precessing systems. Analysis of the observed population of BHNS binaries thus require careful consideration of observational biases and of the probabilistic nature of the characterization of a signal as a BHNS system.

Finally, we note that the availability of reliable GW templates is crucial to the analysis of merger events. In that respect, significant progress have been made in recent years on precessing waveform models~\citep{Schmidt:2012rh,Pan:2013rra,Hannam:2013oca,Smith:2016qas,Khan:2018fmp,Varma:2019csw}, which may be of particular importance for BHNS systems, and on the inclusion of tidal effects in waveform models~\citep{Lackey:2013axa,Bernuzzi:2014owa,Hinderer:2016a,Dietrich:2017aum,Nagar:2018zoe}. Recent high-accuracy numerical simulations of BHNS inspirals~\citep{Foucart:2018inp} show reasonable agreement between tidal models and simulations, except for rapidly spinning neutron stars. It should however be noted that state-of-the art simulations still have numerical errors at the level of $\sim 10\%-20\%$ of the phase difference between BBH and BHNS waveforms, which puts a limit on how far waveform models can be tested in practice.

\section{Merger dynamics}

The merger of a BHNS binary follows one out of two potential pathways: either the neutron star is disrupted by the tidal field of the black hole, leading to mass ejection and the formation of an accretion torus around the black hole; or the neutron star plunges into the black hole whole. Qualitatively, the physical processes leading to these two potential outcomes are well understood~\citep{1976ApJ...210..549L}. As the binary spirals in, the neutron star first reaches either the radius of the innermost stable circular orbit (ISCO) of the black hole $R_{\rm ISCO}$, or the {\it disruption radius} $R_{\rm dis}$. Roughly speaking, disruption happens if $R_{\rm dis} \gtrsim R_{\rm ISCO}$, i.e. if the neutron star is tidally disrupted outside of the ISCO. This division between disrupting and non-disrupting systems creates two classes of events with very distinct observational properties.

Qualitatively, if the neutron star is treated as a test mass and the black hole spin is aligned with the orbital angular momentum of the binary, the ISCO radius scales as $R_{\rm ISCO}=f(\chi_{\rm BH}) GM_{\rm BH}/c^2$, with $f$ a function ranging from $1$ to $9$ and decreasing for increasing (prograde) spins~\citep{1972ApJ...178..347B}. For large mass ratios and in Newtonian physics, the disruption radius scales as $R_{\rm dis}\sim k (M_{\rm BH}/M_{\rm NS})^{1/3} R_{\rm NS}$, with $k$ a numerical constant with a mild dependence on the equation of state and the black hole spin~\citep{1973ApJ...185...43F,Wiggins:1999te}. From these simple scalings, we deduce that disruption will be favored for (a) low-mass black holes; (b) prograde black hole spins; and (c) large neutron star radii.

\begin{figure}[h!]
\begin{center}
\includegraphics[width=0.99\linewidth]{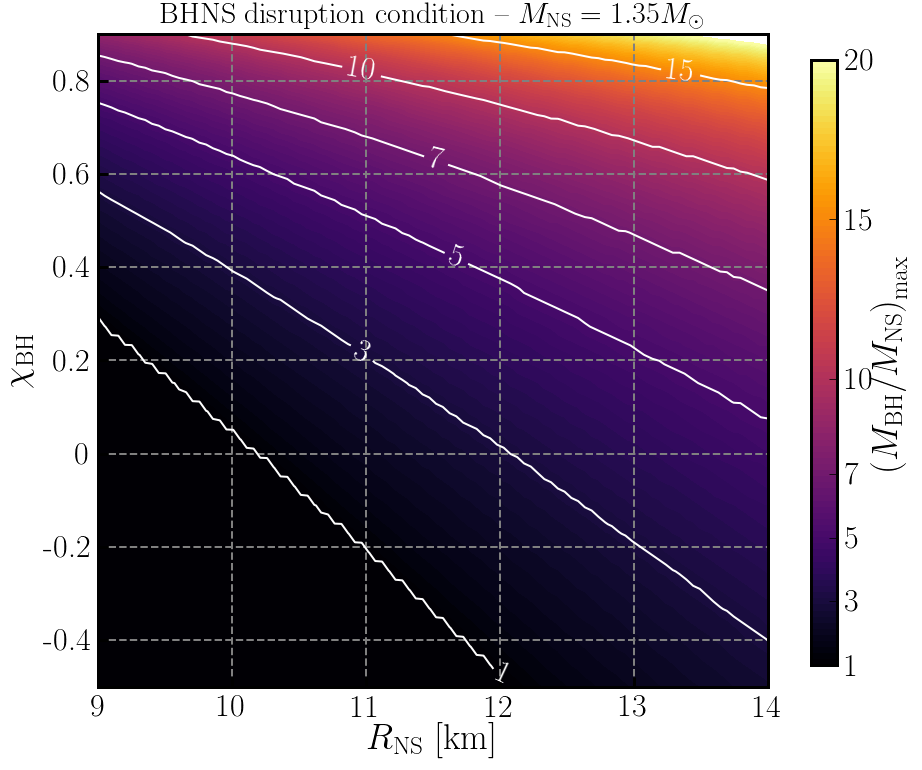}
\end{center}
\caption{Maximum value of the mass ratio $M_{\rm BH}/M_{\rm NS}$ for which a BHNS system will disrupt as a function of the neutron star radius $R_{\rm NS}$ and aligned component of the dimensionless black hole spin $\chi_{\rm BH}$, assuming $M_{\rm NS}=1.35M_\odot$~\citep{Foucart:2018rjc}. Results for other $M_{\rm NS}$ can be obtained by looking at the disruption condition at constant $C_{\rm NS}=GM_{\rm NS}/(R_{\rm NS}c^2)$.}\label{fig:paramspace}
\end{figure}

A more quantitative understanding 
requires general relativistic simulations~\citep{Etienne:2008re,Duez:2008rb,Kyutoku:2010zd,Chawla:2010sw,FoucartEtAl:2011,Kawaguchi:2015}. Simulations tell us that for quasi-circular binaries, {\it mass transfer is always unstable}. We can also {\it predict which systems disrupt} (see Fig.~\ref{fig:paramspace}), and for disrupting systems {\it how much mass remains outside of the black hole after disruption}~\citep{Pannarale:2010vs,Foucart2012,Foucart:2018rjc} -- typically a few tenths of a solar mass. While these predictions were first made for systems with aligned black hole spins, simulations also show that for misaligned black hole spins, simply using in fitting formulae the aligned component of the black hole spin or replacing the ISCO radius by the radius of the innermost stable spherical orbit (ISSO) provides reasonably accurate predictions~\citep{2013PhRvD..87h4053S,Foucart:2013a,Kawaguchi:2015}. Overall, {\it the outcome of a BHNS merger can be predicted with reasonable accuracy as a function of just 3 dimensionless parameters}: the symmetric mass ratio $\eta=M_{\rm NS}M_{\rm BH}/(M_{\rm NS}+M_{\rm BH})^2$, the aligned component of the dimensionless black hole spin $\chi_\parallel$, and the neutron star compaction $C_{\rm NS}$~\citep{Foucart:2018rjc}. However, these models do not apply to systems with large eccentricities: partial disruption of the neutron star is then possible~\citep{Stephens:2011as}, and we do not have reliable predictions for the outcome of the merger in the larger-dimensional parameter space of eccentric BHNS systems. There have also been too few simulations to robustly characterize binaries with rapidly rotating neutron stars. 

For non-disrupting BHNS systems, the merger ends the interesting part of the evolution. The GW signal is practically identical to a BBH system with the same component masses and spins~\citep{Foucart:2013psa,Lackey:2013axa}, there is neither mass ejection nor accretion disk, and we do not expect detectable post-merger EM signals. In the rest of this review, we will thus focus on the more interesting disrupting BHNS systems. However, {\it disrupting systems may very well be a small minority of the observed BHNS binaries}. Even a relatively low mass black hole ($M_{\rm BH}\sim 7M_\odot$) requires a moderate-to-high black hole spin $\chi_\parallel \gtrsim (0.2-0.7)$ to disrupt neutron stars with equations of state compatible with GW170817. The BBH systems detected so far have black holes of high mass and/or low spin~\citep{LIGOScientific:2018mvr,Venumadhav:2019lyq} that would be highly unlikely to disrupt neutron stars -- though the rapidly spinning BH candidate reported in~\cite{Zackay:2019tzo} provides some hope for the existence of disrupting BHNS binaries. While we should be ready for a population of disrupting BHNS mergers, we should acknowledge that {\it the idea that most BHNS mergers undergo tidal disruption is currently disfavored}.

Disrupting BHNS systems provide us with a wealth of additional information. First, the GW signal is cut off when disruption occurs, at a frequency $f_{\rm cut}\sim (1-1.5)\,{\rm kHz}$ that depends on the equation of state of the neutron star. The inclusion of that cut-off frequency in waveform models~\citep{Lackey:2013axa,Pannarale:2015jka} can help constrain the equation of state of neutron stars, complementing the information provided by the tidal dephasing~\citep{Lackey2011,Lackey:2013axa}. Second, a disrupting BHNS binary typically ejects a few percents of a solar mass of material. The ejection of neutron-rich matter at mildly relativistic speeds is extremely important to the study of BHNS and BNS mergers: as the ejecta expands into the surrounding interstellar medium, it undergoes r-process nucleosynthesis, forming many of the heavy elements observed today on Earth. The outcome of the r-process is not, however, unique: more neutron-rich ejecta (approximately, with less than $\sim 25\%$ protons) forms heavier r-process elements than more neutron-poor ejecta~\citep{Wanajo2014,Lippuner2015}. This matters if we wish to understand nucleosynthesis in the Universe, but also to understand the properties of the observable optical/infrared kilonovae powered by radioactive decays in the ejecta. If heavier r-process elements are produced, the opacity of the ejecta increases, causing the kilonova to be dimmer, of longer duration, and shifted from the optical to the infrared~\citep{Kasen:2013xka}. Kilonova signals also contain information about the mass, velocity, composition, and geometry of the ejecta~\citep{2013ApJ...775...18B,Kawaguchi:2016}. Thus, {\it if we can connect the ejecta properties to the parameters of the binary, we can use kilonovae observations to complement and cross-check GW observations of BHNS systems}. 
In BHNS system, the merger ejecta, or {\it dynamical ejecta}, has fairly well constrained properties. It is cold, very neutron-rich ($\sim 5\%$ protons), and moves at an average velocity $v\sim (0.1-0.3)c$. It is also quite different from the dynamical ejecta of BNS mergers: there is more mass ejection in disrupting BHNS binaries, the ejecta is very asymmetric, and there is no neutron poor component to the ejecta that may power an optical kilonova. Fits to the result of numerical simulations have provided us with relatively robust predictions for its mass~\citep{Kawaguchi:2016} and asymptotic velocity~\citep{Kawaguchi:2016,FoucartBhNs2016} that can be used to develop kilonovae models. While higher accuracy predictions for the properties of the dynamical ejecta would certainly be useful in the long term, this phase of the evolution is quite well understood when compared to the formation and evolution of post-merger remnants. 

\section{Post-merger remnants}

In our description of the evolution of BHNS binaries, we have so far only considered the effects of general relativity (GWs, ISCO,...), of ideal hydrodynamics (tides and tidal disruption), and of the nuclear equation of state of cold dense matter in neutrinoless beta-equilibrium. During inspiral and merger, this is generally sufficient to capture the most important observable features of BHNS binaries. This changes dramatically after merger: as bound matter from the disrupted neutron star begins to circularize, mostly through hydrodynamical shocks and interactions between the tidal tail and the forming accretion disk, magnetic fields and neutrinos start to play an important role. Magnetic fields and turbulent eddies will grow due to the Kevin-Helmholtz instability at the disk-tail boundary, heating the disk and driving outflows~\citep{Kiuchi:2015qua}, while neutrinos cool the denser regions of the disk and heat its corona~\citep{Lee:2009uc,Deaton2013,Janiuk:2013lna,FoucartM1:2015}. Neutrino absorption in the corona can drive a disk wind~\citep{Dessart2009} and, more importantly, preferential absorption of electron neutrinos over electron antineutrinos leads to an increase in the ratio of protons to neutrons in the outflows~\citep{FoucartM1:2015}. At later times the growth of the magnetorotational instability leads to an increase in the strength of the magnetic field, angular momentum transport and heating in the disk, accretion of matter onto the black hole, and the production of mildly relativistic outflows for multiple seconds after the merger~\citep{Fernandez2013,Siegel:2017nub,Fernandez:2018kax}. Finally, depending on the large-scale structure of the magnetic field after merger, continuous or more intermittent relativistic jets may be produced $\sim 0.1-1\,{\rm s}$ after merger~\citep{Paschalidis2014,Siegel:2017nub,Ruiz:2018wah,Christie:2019lim}, potentially leading to the production of a SGRB.

Numerical simulations and theoretical models have made important strides in the study of post-merger remnants over the last decade, yet this remains by far the most uncertain part of the evolution. A first problem is that only one magnetohydrodynamics simulation has used sufficient resolution to capture the growth of the Kevin-Helmholtz instability at the disk-tail boundary~\citep{Kiuchi:2015qua}, and it did not include any treatment of the neutrinos. In the absence of cooling, it predicted massive outflows from the forming disk ($50\%$ of the disk mass, an amount comparable to the dynamical ejecta). Lower-resolution simulations including neutrino cooling did not observe significant outflows at this stage~\citep{Deaton2013,FoucartM1:2015}, but lacked the heating provided by the Kevin-Helmholtz instability. The physical answer lies somewhere in between these two extremes, leaving a large uncertainty regarding the mass of hot, mildly relativistic matter that may be ejected during the circularization of the accretion disk. This is particularly problematic because these early post-merger outflows could be the main source of optical kilonovae in BHNS systems.

A second important source of uncertainty is the large scale structure of the magnetic field after merger. Merger simulations have produced jets when a strong dipolar magnetic field was initialized outside of the neutron star before merger~\citep{Paschalidis2014,Ruiz:2018wah}, but no simulation has resolved the growth of a large-scale magnetic field from realistic initial field strengths.
On the other hand, simulations of post-merger remnants show that the large scale structure of the magnetic field has a significant impact on the jet power and the ejected mass~\citep{Christie:2019lim}. This leaves us with important open questions regarding the connection between SGRB properties and the pre-merger characteristics of a BHNS binary, as well as regarding the mechanism for the production of relativistic jets -- e.g. whether a strong magnetic field outside of the neutron star leads to the production of a jet $\sim 100\,{\rm ms}$ after merger~\citep{Paschalidis2014,Ruiz:2018wah}, or a dynamo mechanism within the disk creates a jet later on~\citep{Christie:2019lim}.

One reliable constant in post-merger studies of BHNS systems is that a large fraction $\sim (15-50)\%$ of the bound matter remaining around the black hole after the disruption of a neutron star is ejected in mildly relativistic outflows. There is, however, a wide range of outflow mechanisms. We observe early outflows ($\lesssim 1\,{\rm s}$ post-merger) due to turbulent heating at the disk-tail interface~\citep{Kiuchi:2015qua} and in the inner regions of the disk~\citep{Siegel:2017nub,Fernandez:2018kax}, as well as delayed outflows due to viscous heating and recombination of alpha-particles in the disk~\citep{Fernandez2013,Fernandez:2018kax,Christie:2019lim}. The former have highly uncertain masses, velocities and compositions, in part due to uncertain initial conditions in the post-merger remnant, and in part due to missing physics in the simulations (particularly neutrino radiation transport). The latter are better understood: they are relatively slow ($v\lesssim 0.05c$), and formed of $\sim 20\%-30\%$ protons. 

As post-merger outflows in BHNS mergers have a total mass roughly similar to that of the dynamical ejecta, they have a large impact on the properties of BHNS-powered kilonovae. Uncertainties in their mass (by a factor of 2-3), velocity, and composition (neutron rich/neutron poor) are the main source of error in the construction of theoretical models of BHNS outflows today. A better understanding of the post-merger winds, along with better nuclear models and improved understanding of the heating rate of merger outflows~\citep{Barnes:2016}, are critical to the production of reliable kilonova models for BHNS binaries. Currently, models either ignore the post-merger ejecta~\citep{Kawaguchi:2016}, take only some of the post-merger outflows into account~\citep{Barbieri:2019kli}, or suffer from large uncertainties due to our lack of understanding of the post-merger ejecta~\citep{Andreoni:2019qgh,Coughlin:2019zqi}.

Finally, let us comment briefly on our understanding of BHNS binaries as engines for SGRBs. Relativistic jets have now been produced in simulations of BHNS merger~\citep{Paschalidis2014,Ruiz:2018wah}\footnote{`Jets' here are outflow regions with large Poynting flux, that cannot reach Lorentz factor of more than a few due to the limits of existing merger simulations.}  and/or of their post-merger remnant disks~\citep{Siegel:2017nub,Christie:2019lim}. We also know that the properties of the jet depend on the large scale structure of the post-merger magnetic field. However, connecting that large scale structure to the pre-merger properties of the system remains an important unsolved problem. It is unclear whether observations or simulations will first constrain the magnetic field structure of the remnant of a BHNS merger. At the moment, the most reliable information that comes from the joint observation of a SGRB and GW signal from a BHNS binary is that the neutron star was disrupted. 

\section{Discussion}

With the advent of GW astronomy, the study of BHNS mergers is undergoing a rapid transformation. More efforts are now being directed towards the modeling and interpretation of multi-messenger observations of binary mergers. It has also become clear that the study of BHNS systems suffers from significant complications when compared with BNS systems: statistical uncertainties in the individual masses of the merging compact objects make it difficult to unequivocally characterize a GW event as a BHNS binary, and many BHNS binaries likely involve high-mass and/or low-spin black holes for which the neutron star plunges whole into the black hole, preventing the emission of detectable post-merger EM signals. 

To make optimal use of the available observational data, reliable models of the GW and EM signals powered by BHNS binaries are required. On the GW side, a number of models including finite-size effects for BHNS and/or BNS systems have been developed~\citep{Lackey:2013axa,Bernuzzi:2014owa,Hinderer:2016a,Dietrich:2017aum,Nagar:2018zoe}. Before merger, these models agree with numerical simulations of BHNS binaries within current numerical errors, except in the most extreme cases tested so far~\citep{Foucart:2018inp}. Models for the impact on the GW signal of the disruption of the neutron star have more room to improve: they remain only order-of-magnitude accurate, and typically limited in their coverage of the BHNS parameter space~\citep{Lackey:2013axa,Pannarale:2015jka}. 

On the EM side, a first model combining information from SGRBs and kilonovae was recently developed by~\cite{Barbieri:2019kli}, adding to a previously developed model for the kilonova signals powered by the dynamical ejecta of BHNS mergers~\citep{Kawaguchi:2016}. Existing models remain however limited by our lack of understanding of post-merger outflows, as well as by nuclear physics and radiation transport uncertainties~\citep{Barnes:2016}. In particular, the large scale structure of magnetic fields within and outside of the post-merger accretion disk is not, at this point, well constrained by merger simulations, despite its large impact on post-merger outflows and on the properties of SGRBs~\citep{Christie:2019lim}. To make optimal use of upcoming multi-messenger observations (or even non-detections), it is thus important to develop improved kilonova and SGRB models, and properly characterize model uncertainties.

\section{Funding}
F.F. gratefully acknowledges support from the DOE through Early Career award DE-SC0020435, from NASA through grant 80NSSC18K0565, and from the NSF through grant PHY-1806278.

\bibliography{References}

\end{document}